\documentclass[11pt,english]{article}
\usepackage[T1]{fontenc}
\usepackage[latin9]{inputenc}
\usepackage{listings}
\usepackage{geometry}
\usepackage{amssymb}
\usepackage{graphicx}
\usepackage{babel}
\begin{document}

\title{A String-Based Public Key Cryptosystem}

\author{M. Andrecut}

\maketitle
\bigskip{}

{

\centering Unlimited Analytics Inc.

\centering Calgary, AB, Canada

\centering mircea.andrecut@gmail.com

}

\bigskip{}

\bigskip{}

\begin{abstract}
Traditional methods in public key cryptography are based on number theory, 
and suffer from problems such as dealing with very large numbers, 
making key creation cumbersome. Here, we propose
a new public key cryptosystem based on strings only, which avoids
the difficulties of the traditional number theory approach. The security
mechanism for public and secret keys generation is ensured by a recursive
encoding mechanism embedded in a quasi-commutative-random function,
resulted from the composition of a quasi-commutative function with
a pseudo-random function. In this revised version of the paper we show that 
the eavesdropper's problem of the proposed cryptosystem has a solution, 
and we give the details of the solution.
 
\end{abstract}

\bigskip{}

\section{Introduction}

In a symmetrical key cryptosystem \cite{key-1}, such as AES (Advanced Encryption Standard), 
two users Alice and
Bob must first agree on a common secret key. If Alice communicates
the secret key to Bob, a third party, Eve, might intercept the key,
and decrypt the messages. In order to avoid such a situation Alice and Bob can
use an asymmetric public key cryptosystem \cite{key-1}, which provides a secure mechanism
to exchange information between two users. 

In public key cryptography each user has a pair of cryptographic keys,
consisting of a public key and a secret private key. These keys are 
related through a hard mathematical inversion problem, such that the 
private key cannot be practically derived from the public key. 

The two main directions of public key cryptography
are the public key encryption and the digital signatures. Public key
encryption is used to ensure confidentiality. In this case, the data
encrypted with the public key can only be decrypted with the corresponding
private key, and vice versa. Digital signatures are used to ensure
authenticity. In this case, any message signed
with a user's private key can be verified by anyone having the user's public key, 
proving the authenticity of the message. 

A standard implementation of public key cryptography is based on the
Diffie-Hellman (DH) key agreement protocol \cite{key-2}. The protocol allows
two users to exchange a secret key over an insecure communication
channel. The platform of the DH protocol is the multiplicative group
$\mathbb{Z}_{p}$ of integers modulo a prime $p$. The DH protocol
can be described as following: 
\begin{enumerate}
\item Alice and Bob agree upon the public integer $g\in\mathbb{Z}_{p}$.
\item Alice chooses the secret integer $a$.
\item Alice computes $A=g^{a}\,\mathrm{mod}\, p$, and publishes $A$.
\item Bob chooses the secret integer $b$, 
\item Bob computes $B=g^{b}\mathrm{\, mod}\, p$, and publishes $B$.
\item Alice computes the secret integer $K_{A}=B^{a}\mathrm{\, mod\,}p=g^{ba}\,\mathrm{mod\,}p$.
\item Bob computes the secret integer $K_{B}=A^{b}\mathrm{\, mod\,}p=g^{ab}\mathrm{\, mod\,}p$.
\end{enumerate}
It is obvious that both Alice and Bob calculate the same integer $K=K_{A}=K_{B}$,
which then can be used as a secret shared key for symmetric encryption,
improving the performance of the communication channel, since symmetric-key
algorithms are generally much faster.

Assuming that the eavesdropper Eve knows $p,g,A$ and $B$, she needs
to compute the secret key $K$, that is to solve the discrete logarithm
problem: 
\begin{equation}
A=g^{a}\mathrm{\, mod}\, p,
\end{equation}
for the unknown $a$. If $p$ is a very large prime of at least 300
digits, and $a$ and $b$ are at least 100 digits long, then the problem
becomes computationally hard (exponential time in $\log\, p$), and
it is considered infeasible. For maximum security $p$ should be a safeprime, 
i.e. $(p-1)/2$ is also a prime, and $g$ a primitive root of $p$ \cite{key-3}. 

An important short coming of traditional public key cryptosystems
is the computation with very large numbers and prime numbers (300 digit),
which is difficult and not very efficient. For similar key sizes, the DH method has a similar 
key strength as the mathematically related RSA method \cite{key-4}, which
is based on the hard integer factorization problem. Currently, no
classical algorithm that can factor large numbers efficiently is known,
which makes the RSA and the DH methods, widely used cryptographic
protocols. However, this problem can be solved on a quantum computer
in polynomial time using Shor's algorithm \cite{key-5}. Therefore, the current public key cryptographic 
protocols are insecure if sufficiently large quantum computers become available, putting many sensitive systems 
like e-commerce and Internet banking at risk. This is why there is an ongoing search for other
methods, where similar key exchange mechanisms can be implemented
more efficiently, and are immune to quantum algorithms attacks.

Here, we propose a new public key cryptosystem based on strings only,
which avoids the traditional number theory approach. The security
of the proposed cryptosystem is ensured by a recursive encoding mechanism
embedded in a quasi-commutative-random function, which is a composition
of a quasi-commutative function with a pseudo-random function. 
Also, in this revised version of the paper we show that the eavesdropper's problem 
of the proposed cryptosystem has a solution based on the modular multiplicative 
inverse, and we give the details of the solution.

\section{Quasi-commutative transformation}

Let us consider the following set:
\begin{equation}
\Omega=\{0,1,2,...,p-1|p=2n,n>1\},
\end{equation}
where the total number of elements is even. For example, if the intention
is to build a crytosystem based on strings where the characters $\alpha\in\Omega$
are representable on a byte word (8 bits), then one can consider $p=2^{8}=256$.
However, we should emphasize that this choice is not a restriction,
and the results obtained here are valid for any even integer $p=2n$.
Thus, we assume that any element: 
\begin{equation}
s=\{s_{0},s_{1},...,s_{N-1}\}\in\Omega^{N}
\end{equation}
is a string of length $N$ defined over the set $\Omega$. 

A function $F:\Omega\times\Omega\rightarrow\Omega$ is said to be
\emph{quasi-commutative} if for any $\xi,\alpha,\beta\in\Omega$ we
have \cite{key-6}: 
\begin{equation}
F(F(\xi,\alpha),\beta)=F(F(\xi,\beta),\alpha).
\end{equation}
For example, the function used in both DH and RSA algorithms: 
\begin{equation}
F_{p}(\xi,\alpha)=\xi^{\alpha}\,\mathrm{mod}\, p,
\end{equation}
is quasi-commutative, because:
\begin{equation}
F_{p}(F_{p}(\xi,\alpha),\beta)=\xi^{\alpha\beta}\,\mathrm{mod}\, p=F_{p}(F_{p}(\xi,\beta),\alpha).
\end{equation}

Here we consider the function $G_{w}:\Omega\times\Omega\rightarrow\Omega$
defined as:
\begin{equation}
G_{w}(\xi,\alpha)=[(w\alpha+1)\xi+\alpha]\,\mathrm{mod}\, p,\quad w\in\mathbb{N}.
\end{equation}
The function $G_{w}$ is quasi-commutative, since we have:
\[
G_{w}(G_{w}(\xi,\alpha),\beta)=(w\beta+1)[(w\alpha+1)\xi+\alpha]+\beta=
\]
\[
=(w\alpha+1)(w\beta+1)\xi+\alpha(w\beta+1)+\beta=
\]
\[
=(w\alpha+1)(w\beta+1)\xi+\beta(w\alpha+1)+\alpha=
\]
\begin{equation}
=(w\alpha+1)[(w\beta+1)\xi+\beta]+\alpha=G_{w}(G_{w}(\xi,\beta),\alpha),
\end{equation}
where we have omitted $\mathrm{mod}\, p$, in order to simplify the
notation.

The quasi-commutative property ensures that if one starts with an
initial value $\xi\in\Omega$ and a set of values $\{\alpha_{0},\alpha_{1},...,\alpha_{N-1}\}\in\Omega^{N}$,
then the result of the composition:
\begin{equation}
r=G_{w}(...(G_{w}(G_{w}(\xi,\alpha_{0}),\alpha_{1}),...,\alpha_{N-1})
\end{equation}
would not change if the order of the values $\alpha_{n}$ were permuted \cite{key-6}. 

We should note that if the parameter $w$ is even, $w=2q$, $q\in\mathbb{N}$,
then the function $G_{w}$ does not have fixed points.\textbf{\textit{
}}In order to prove this, let us consider the fixed point equation:

\begin{equation}
G_{w}(\xi,\alpha)=[(w\alpha+1)\xi+\alpha]\,\mathrm{mod}\, p=\xi,
\end{equation}
which is equivalent to
\begin{equation}
(w\alpha+1)\xi+\alpha=\xi+\alpha mp,
\end{equation}
and 
\begin{equation}
w\xi=mp-1.
\end{equation}
Since $mp-1$ is odd, $w$ must be also odd in order to have a solution
of the above equation. For example, if $w=1$ then $\xi=(mp-1)\,\mathrm{mod}\, p=p-1$
is a solution of the fixed point equation. Similarly, if $w=3$, then
$\xi=p-3$ is a solution of the fixed point equation. However, if
$w$ is even, then the fixed point equation has no integer solutions.
Thus, the parameter $w$ in the function $G_{w}$ must be even, $w=2q$,
$q\in\mathbb{N}$, in order to avoid a fixed point in repeated iterations.

\section{Pseudo-random transformation}

Let us now consider the following problem. Given a string ``seed'':
\begin{equation}
s^{(0)}=\{s_{0}^{(0)},s_{1}^{(0)},...,s_{N-1}^{(0)}\}\in\Omega^{N},
\end{equation}
find a transformation $R$, that generates a non-repeating and pseudo-random
sequence of strings, through a recursive application:
\begin{equation}
s^{(t+1)}=R(s^{(t)})=...=R^{(t+1)}(s^{(0)})=\underbrace{R(R(...R(s^{(0)})))}_{t+1}.
\end{equation}

The transformation $R$ can be for example a cryptographic hash function,
which is considered impossible to invert, a random number generator,
or a chaotic deterministic function. Here we prefer to use a secure
cryptographic hash function, due to its ability to deal directly
with input strings of any length. However, following a similar approach, 
one can also adapt a cryptographically secure random number generator
(for example AES), or a chaotic deterministic function. 

An ideal cryptographic hash function has the following properties:
(a) it computes a hash value for any given string; (b) it is impossible
to invert, i.e. to generate a string that has a given hash; (c) any
change in the input string triggers a change in the hash value; (d)
it is impossible to find two different strings with the same hash
value. Therefore, a cryptographic hash function should provide a random
transformation of the input string. Typical hash functions that can
be used for the role of the $R$ transformation are the standard SHA-1,
SHA-2 and SHA-3 cryptographic hash functions, published by the National
Institute of Standards and Technology (NIST) as a U.S. Federal Information
Processing Standard (FIPS). 

Most cryptographic hash functions are designed to take a string of
any length as input and produce a fixed-length hash string value.
Here, we assume that the computed output hash string has a fixed length
$L>0$, such that:
\begin{equation}
R:\Omega^{N}\rightarrow\Omega^{L},\:\forall N\geq0.
\end{equation}

\section{The proposed cryptosystem}

For any element $\xi\in\Omega$ and string $s=\{s_{0},s_{1},...,s_{N-1}\}\in\Omega^{N}$
we define the following recursive transformation:
\begin{equation}
T:\Omega^{N+1}\rightarrow\Omega,
\end{equation}
where
\begin{equation}
T(\xi,s)=G_{w}(...(G_{w}(G_{w}(\xi,s_{0}),s_{1}),...,s_{N}).
\end{equation}
According to the results from the previous sections, this transformation
satisfies the quasi-commutative property. Also, we should note that
from the algorithmic point of view, the transformation $T$ can be
calculated as following: 
\begin{equation}
T(\xi,s)=\left\{ \begin{array}{l}
\tilde{\xi}\leftarrow\xi\\
\mathrm{for}\: n=0...N-1\\
\quad\tilde{\xi}\leftarrow G_{w}(\tilde{\xi},s_{n})\\
\mathrm{return}\:\tilde{\xi}
\end{array}\right..
\end{equation}

Let us now define a more complex transformation $W$, as a composition
of $T$ and $R$. For any two strings $x=\{x_{0},x_{1},...,x_{K-1}\}\in\Omega^{K}$
and $s=\{s_{0},s_{1},...,s_{N-1}\}\in\Omega^{N}$ we define:
\begin{equation}
W:\Omega^{K+N}\rightarrow\Omega^{K},
\end{equation}
with the components $W_{k}$, $k=0,1,...,K-1$, given by:

\begin{equation}
W_{k}=T(x_{k},R^{(k+1)}(s)),
\end{equation}
satisfying the quasi-commutative property, since $T$ is quasi-commutative.
The whole transformation is also quasi-commutative, and algorithmically
it can be computed as following:
\begin{equation}
W(x,s)=\left\{ \begin{array}{l}
\tilde{s}\leftarrow R(s)\\
\tilde{x}_{0}\leftarrow T(x_{0},\tilde{s})\\
\mathrm{for}\: k=1...K-1\\
\quad\tilde{s}\leftarrow R(\tilde{s})\\
\quad\tilde{x}_{k}\leftarrow T(x_{k},\tilde{s})\\
\mathrm{return}\:\tilde{x}
\end{array}\right..
\end{equation}
An important aspect of the $W$ function, is that each component $x_{k}$
of the input string $x$ is encoded using a different string $\tilde{s}$
obtained by recursively applying the pseudo-random cryptographic hash function
$R$ to the secret string $s$. 

Since now all the necessary ingredients have been defined, the proposed
key generation and exchange protocol can be formulated as following:
\begin{enumerate}
\item Alice and Bob agree upon the public string $g=\{g_{0},g_{1},...,g_{K-1}\}\in\Omega^{K}$. 
\item Alice chooses the secret string $a=\{a_{0},a_{1},...,a_{N}\}\in\Omega^{N}$.
\item Alice computes the string $A=W(g,a)\in\Omega^{K}$, and publishes
$A$. 
\item Bob chooses the secret string $b=\{b_{0},b_{1},...,b_{M}\}\in\Omega^{M}$. 
\item Bob computes the string $B=W(g,b)\in\Omega^{K}$, and publishes $B$.
\item Alice calculates the secret string $s_{A}=W(B,a)=W(W(g,b),a)\in\Omega^{K}$.
\item Bob calculates the secret string $s_{B}=W(A,b)=W(W(g,a),b)\in\Omega^{K}$.
\end{enumerate}
Both Alice and Bob obtain the same secret key $s=s_{A}=s_{B}$,
since $W$ satisfies the quasi-commutativity property:
\begin{equation}
W(W(g,b),a)=W(W(g,a),b).
\end{equation}
The flow chart showing the above key generation and exchange protocol
is given in Figure 1. Also, the C code implementing the proposed key 
generation and exchange protocol, and a typical key exchange simulation
example, are given in the Appendix.

\section{The hard problem}

Assuming that the eavesdropper Eve knows $g,A$ and $B$ the hard
problem is to compute the strings $a\in\Omega^{N}$ and respectively
$b\in\Omega^{M}$. This is a hard problem, since even the length ($N$ and $M$) of
the strings is kept secret by Alice and Bob, and the
security is ensured by the above described recursive encoding mechanism
embedded in the quasi-commutative-random function $W$, which is a
composition of a quasi-commutative function $T$ with a pseudo-random
function $R$. 

The encoding mechanism can also be described with the following recursive
equations for the evolution of $a$ and $A_{k}$, $k=0,1,...,K-1$:
\begin{equation}
a^{(k)}=R^{(k+1)}(a)=\underbrace{R(R(...R(a)))}_{k+1},
\end{equation}
and respectively
\begin{equation}
A_{k}=[g_{k}\prod_{n=0}^{L-1}(wa_{n}^{(k)}+1)+\sum_{i=0}^{L-2}a_{i}^{(k)}\prod_{n=i+1}^{L-1}(wa_{n}^{(k)}+1)+a_{N_{k}-1}^{(k)}]\mathrm{\, mod}\, p,
\end{equation}
with the unknowns $a_{0}^{(k)},a_{1}^{(k)},...,a_{L-1}^{(k)}$. We
should note that for each $k=0,1,...,K-1$, any permutation $\tilde{a}^{(k)}$
of $a^{(k)}$ is also a solution of the equation (24). Assuming that
by chance Eve finds a solution $\tilde{a}^{(0)}$ for the first equation,
$k=0$, which is in fact a permutation of a hash of $a$, then one
may think that she may use the recursion equation (23) to find $\tilde{a}^{(k)}$.
This is not possible either, because the results $\tilde{a}^{(k)}$
obtained via $R^{(k+1)}$ are completely dependent on the order of
the characters in the string $\tilde{a}^{(0)}$. Therefore, any permutation
$\tilde{a}^{(0)}$ of $a^{(0)}$ will give a different result $\tilde{a}^{(k)}$.
The recursion $R^{(k+1)}$ gives the correct result if and only if
$\tilde{a}^{(0)}\equiv a^{(0)}$.  

\section{Solution of the eavesdropper's problem}

We assume that Eve knows the public information $A$ and $B$, 
and we show that she can use this information to find the shared
secret key $s\equiv s_{A}=s_{B}$. Since $A=W(g,a)\in\Omega^{K}$,
for $k=0,1,...,K-1$ we consider:
\begin{equation}
A_{k}=[(wg_{k}+1)e_{k}+g_{k}]\mathrm{\, mod}\, p
\end{equation}
where the quantity $e=\{e_{0},e_{1},...,e_{K-1}\}$ will act as Alice's
private key $a$. From here we obtain: 
\begin{equation}
A_{k}-g_{k}=(wg_{k}+1)e_{k}\mathrm{\, mod}\, p,
\end{equation}
\begin{equation}
e_{k}=(wg_{k}+1)^{-1}(A_{k}-g_{k})\mathrm{\, mod}\, p,
\end{equation}
The calculation of the modular inverse is shown in the program from the 
Appendix. We also have $s=W(B,a)\in\Omega^{K}$, and by replacing
$a$ with $e$ we obtain the shared secret key: 
\begin{equation}
s_{k}=[(wB_{k}+1)e_{k}+B_{k}]\mathrm{\, mod}\, p,\quad k=0,1,...,K-1.
\end{equation}

\section{Conclusion}

In conclusion, we have presented a new public key cryptosystem based
on strings, which avoids the cumbersome key generation using the traditional
number theory approach. The security mechanism for public and secret
keys generation is ensured by a recursive encoding mechanism embedded
in a quasi-commutative-random function, resulted from the composition
of a quasi-commutative function with a pseudo-random function. 
Also, In this revised version of the paper we have shown that the Eavesdropper's problem 
of the proposed cryptosystem has a solution based on the modular multiplicative 
inverse, and we have given the details of the solution.

\section*{Acknowledgement}
The author would like to thank C. Monico for his comments.

\begin{figure}
\centering{}\includegraphics[clip,scale=0.9]{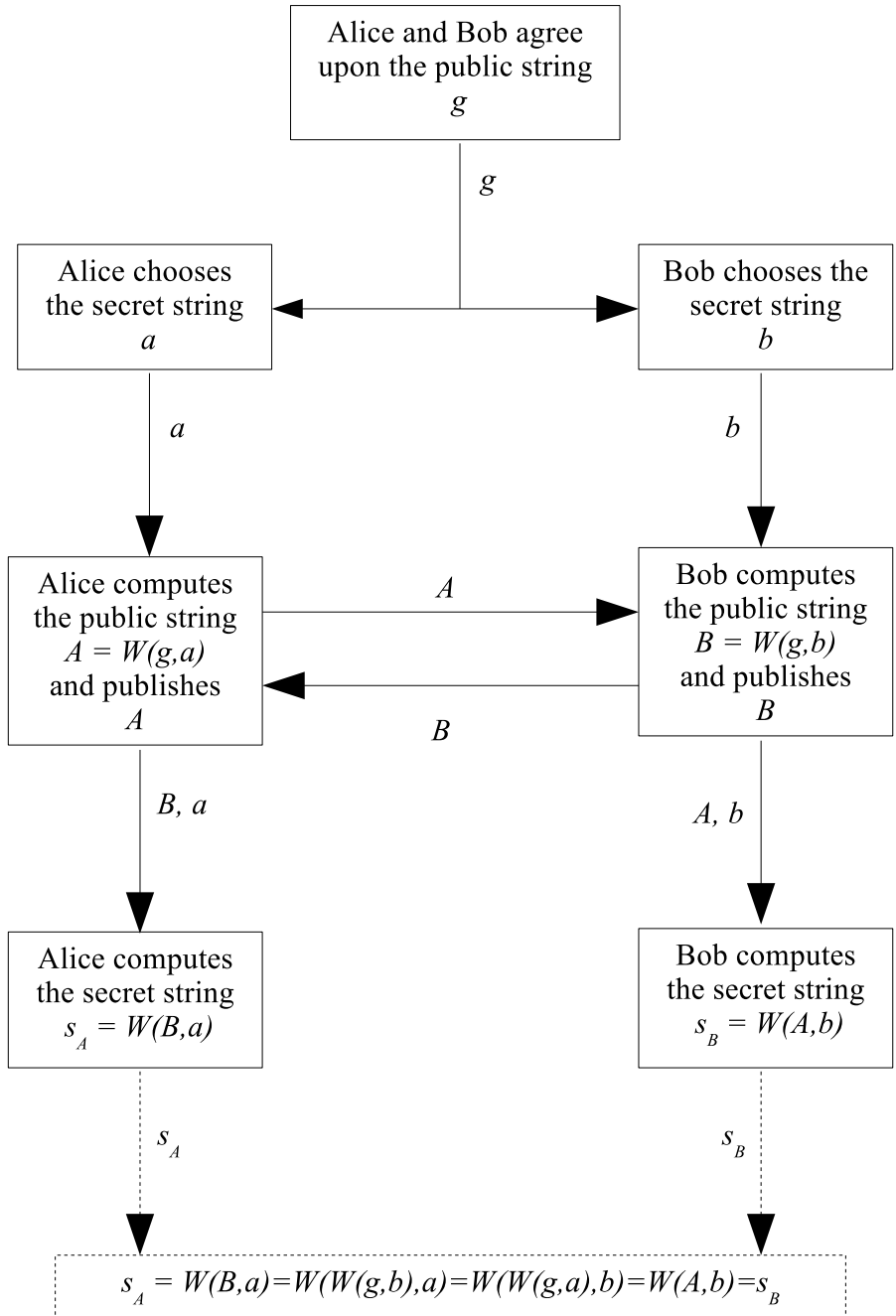}\caption{The flow chart of the key generation and exchange protocol.}
\end{figure}

\section*{Appendix}

Below we give the C code of the proposed cryptosystem and the solution of the eavesdropper's problem.  

\begin{lstlisting}[basicstyle={\small\ttfamily},breaklines=true,tabsize=2]

//keys.c, C implementation of the key exchange protocol
#include <stdlib.h>
#include <stdio.h>
#include <time.h>
#include <openssl/sha.h>

unsigned char *W(unsigned char *x, unsigned K,
                 unsigned char *s, unsigned N){
	unsigned k, n, w = 2;
	unsigned char *y = malloc(K);

	unsigned char *z = SHA512(s, N, 0);
	y[0] = x[0];
	for(n=0; n<SHA512_DIGEST_LENGTH; n++){
		y[0] = (w*z[n] + 1)*y[0] + z[n];
	}
	for(k=1; k<K; k++){
		z = SHA512(z, SHA512_DIGEST_LENGTH, 0);
		y[k] = x[k];
		for(n=0; n<SHA512_DIGEST_LENGTH; n++){
			y[k] = (w*z[n] + 1)*y[k] + z[n];
		}
	}
	return y;
}

void sprint(unsigned char *x, unsigned N){
	unsigned n;

	for(n=0; n<N; n++){
		if(n % 32 == 0) printf("\n");
		else printf("%02x", x[n]);
	}
	printf("\n");
}

/*****************************************************/
/* Code added to solve Eve's problem. */

unsigned char inverseMod256(unsigned char x){
  static int initialized=0;
  static unsigned inv[256];
  unsigned int i, j;

  if (!(initialized)){
    /* Build a table of inverses. */
    for(i=0; i<256; i++) inv[i] = 0;
    for(i=0; i<256; i++){
      for(j=0; j<256; j++){
        if((i*j)%256 == 1){
          inv[i] = (unsigned char)j;
          inv[j] = (unsigned char)i;
        }
      }
    }
    initialized=1;
  }
  return inv[x];
}

void Eve(unsigned char *g, unsigned char *A, 
         unsigned char *B, unsigned K, unsigned w){

  unsigned char *e = malloc(K), *Se = malloc(K);
  int j;

  for(j=0; j<K; j++){
  	e[j] = (A[j] - g[j])*inverseMod256(w*g[j] + 1);
  	}
  /* e[0],...,e[K-1] acts like Alice's private key.
     Now find the shared secret key. */
  for(j=0; j<K; j++){ 
  	Se[j] = w*B[j]*e[j] + B[j] + e[j];
  	}
  printf("\n\nEve has recovered the shared secret key:\n");
  sprint(Se, K);
}
/*****************************************************/


main(int argc, char *argv[]){
	unsigned N = 253, M = 121, K = 127;
	unsigned n, m, k;
	srand((unsigned char) time(0) + getpid());

	unsigned char *g = malloc(K);
	for(k=0; k<K; k++) g[k] = rand() % 256;
	printf("\nGenerator: public g"); sprint(g, K);

	unsigned char *a = malloc(N);
	for(n=0; n<N; n++) a[n] = rand() % 256;
	printf("\nAlice: secret key a"); sprint(a, N);

	unsigned char *b = malloc(M);
	for(m=0; m<M; m++) b[m] = rand() % 256;
	printf("\nBob: secret key b"); sprint(b, M);

	unsigned char *A = W(g, K, a, N);
	printf("\nAlice: public key A"); sprint(A, K);

	unsigned char *B = W(g, K, b, M);
	printf("\nBob: public key B"); sprint(B, K);

	unsigned char *Sa = W(B, K, a, N);
	printf("\nAlice: shared secret key s = Sa"); sprint(Sa, K);

	unsigned char *Sb = W(A, K, b, M);
	printf("\nBob: shared secret key s = Sb"); sprint(Sb, K);

	/* Now, Eve can use the public information A and B
       to find the shared secret key. */
	
	Eve(g, A, B, K, 2);

	free(a); free(b); free(g); free(A);
	free(B); free(Sa); free(Sb);
	return 0;
}
\end{lstlisting}

The code example (keys.c) has been tested on a LINUX PC, running Ubuntu
14.04 LTS, using the standard GCC compiler and OpenSSL implementation
of the Secure Socket Layer (SSL), and related cryptographic tools.
In this example, the role of the secure cryptographic hash function
$R$ is taken by the SHA-2 function which provides hash strings with
a length of 64 bytes (512 bits), and we set $p=256$. 

The required compilation and run steps are: 
\begin{lstlisting}[tabsize=2]
	gcc -lssl -lcrypto keys.c -o keys
	./keys
\end{lstlisting}

For illustration purposes, the results obtained for one instance run
are given below in hexadecimal. One can see that both Alice and Bob
compute the same secret key (Sa = Sb), using completely different
secret and public keys. One can can also see that Eve can recover 
the secret key using the public information $A$ and $B$. 

\begin{lstlisting}[basicstyle={\small\ttfamily}]

Simulation results:

Generator: public g
93444ff1380add6ae67dba5444e16cffa02679ba50e6c66cf72b18c7cf53d3
972253d02c303a12aef467f2d6d3f276f96b304951f6b64922ce10f121e353
06a68932d6c34584b8ac778e7f690478d434c2252a786e4c467e3d67629020
36a99a0d6ddf91258b08b30a71b78345eb456b15bdd96203589f6aba308b

Alice: secret key a
6634bc73a19c05c6270d79327f30b5c41bfa2f31b70993ba6132241b62af3d
e4f93c8595414cbd4fc5efcef6a492119ec24255cbd50f2c0833476ae38433
7d704c13b198d0005ebfce546361650123a856ee7d651a859861f07be52342
938f7545274545850414d967753f6898e7be866423a0eabc01da37e6fd7a48
09bed630031bb6082f8f6fa4ced83cb596c21aba62047663dead49db27926c
504261535e175b8da6cb3275a36e2a393144f3934869f7261740023ed26e6f
b1d0760fe7d19c8d9cce023f3d2d796e716c01bad6f8e0ed39e22b0b519a2e
6aa4115175addf127ce151b90eca278037283a0d211afa5afd25654e

Bob: secret key b
93502a37617cad0e5bbf8a3c10434bdb6acb1293051fb41f190e1c3e736afe
ba293e1ba5eb2a00aab43cbbf887966252a8f557c7a977e0b7931e2bfe1d32
4670d4eb5cfeeb06b227c1aaaf570d01ff0259c6acd0a66363c58e61e2c01a
31ee138decfe939e255549d4ac56d6ac582f7204ff196862

Alice: public key A
5e47a84e9c3934d60b9c86eac559446036b4c3922eb5d69f6214d0ed571399
55e5f2a66019720222fd1c280a811f54fceca2c1f66d532cd35c0ed5a01c0f
0f7626a0f3981a163a79ed4caf92e2e992121a4ef84708f4a510555a140f5f
48f78dae89da8100ccb46a4140ceb1fd03cf382d99532c233a060c065985

Bob: public key B
64f5bad9a33bc6fb07cf101e6ec86507d607daa28dec4878c47573d23b9bb4
76d04a6a0c37b400ab8f364b048b8fad94723ed81a0437c3fcc3373842e93d
d0ad6743cce4138fcd7afe363ead967c6696b979dea48a65aa5a04d47d0133
1761eda41e9eeb6f45196e74d37bccc7c3b127898047dcf1440f3527e7f5

Alice: shared secret key s = Sa
c96eed661ff0575fdec2acaca370ed58b4a1c0fab73f98b3b1cafb74035b8e
7097c7b8409ecc30078e551da81906ffed952800dd0fb65a918d19442b16c9
dd5d081d15ffcc4de31340d4ceeeecf51caca15a9c1b945d219c0c794bde4c
dd0f6aefe28f1b9e321dbb07f6dade1f5b8b9ce154898e91d656134b6e27

Bob: shared secret key s = Sb
c96eed661ff0575fdec2acaca370ed58b4a1c0fab73f98b3b1cafb74035b8e
7097c7b8409ecc30078e551da81906ffed952800dd0fb65a918d19442b16c9
dd5d081d15ffcc4de31340d4ceeeecf51caca15a9c1b945d219c0c794bde4c
dd0f6aefe28f1b9e321dbb07f6dade1f5b8b9ce154898e91d656134b6e27

Eve has recovered the shared secret key:
c96eed661ff0575fdec2acaca370ed58b4a1c0fab73f98b3b1cafb74035b8e
7097c7b8409ecc30078e551da81906ffed952800dd0fb65a918d19442b16c9
dd5d081d15ffcc4de31340d4ceeeecf51caca15a9c1b945d219c0c794bde4c
dd0f6aefe28f1b9e321dbb07f6dade1f5b8b9ce154898e91d656134b6e27
\end{lstlisting}


\begin{thebibliography}{1}
\bibitem{key-1}Douglas R. Stinson, Cryptography Theory and Practice,
third edition, CRC Press, Boca Raton, Florida, 2006.

\bibitem{key-2}W. Diffie, M. Hellman, New directions in cryptography,
IEEE Transactions on Information Theory 22 (6): 644\textendash{}654
(1976). 

\bibitem{key-3}Randall K. Nichols, ICSA guide to Cryptography,
McGraw-Hill, New York, 1999.

\bibitem{key-4}R. Rivest, A. Shamir, L. Adleman, A Method for Obtaining
Digital Signatures and Public-Key Cryptosystems, Communications of
the ACM 21 (2): 120\textendash{}126 (1978).

\bibitem{key-5}P. W. Shor, Algorithms for quantum computation: Discrete
logarithms and factoring, Proc. 35nd Annual Symposium on Foundations
of Computer Science (Shafi Goldwasser, ed.), IEEE Computer Society
Press, 124-134 (1994). 

\bibitem{key-6}J. Benaloh, M. de Mare, One-way accumulators: a decentralized alternative to digital signatures, 
Advances in Cryptology-Eurocrypt'93, LNCS, vol. 765, Springer-Verlag, 274-285, 1993.


\end{thebibliography}
\end{document}